\documentclass[]{article}

\usepackage{graphicx}
\newcommand{\bra}[1]{\langle #1 |}
\newcommand{\ket}[1]{| #1 \rangle}
\newcommand{\brahket}[3]{\langle #1 | #2 | #3 \rangle}
\newcommand{\braket}[2]{\langle #1 | #2 \rangle}
\newcommand{\ketbra}[2]{\ket{#1}\bra{#2}}

\newcommand{\mydt}[1]{\frac{\partial #1}{\partial t}}
\newcommand{\mysum}[2]{\sum\limits_{#1}^{#2}}
\newcommand{\myint}[2]{\int\limits_{#1}^{#2}}

\newcommand{\mybpar}[1]{\left( #1 \right)}
\newcommand{\beq}{\begin{equation}}
\newcommand{\eeq}{\end{equation}}
\newcommand{\beqa}{\begin{eqnarray}}
\newcommand{\eeqa}{\end{eqnarray}}
\newcommand{\mymat}[2]{\left( \begin{array}{*{#1}{c}} #2 \end{array} \right) }
\newcommand{\myvec}[1]{\left( \begin{array}{*{1}{c}} #1 \end{array} \right) }

\newcommand{\mypsi}[1]{\ket{\psi_{#1}}}

\begin{document}
\thispagestyle{plain}
\pagestyle{plain}

\title{Non Equilibrium Green's Functions for Dummies: \\ 
Introduction to the One Particle NEGF equations} 
\author{Magnus Paulsson\thanks{mpn@mic.dtu.dk}\\
Dept. of micro- and nano-technology, NanoDTU,\\Technical University of Denmark}

\date{\today} 
\maketitle

\begin{abstract}
Non equilibrium Green's function methods are regularly used to calculate current and charge densities
in nanoscale (both molecular and semiconductor) conductors under bias. This method is mainly used for 
ballistic conduction but may be extended to include inelastic scattering. 
In this tutorial paper the NEGF equations for the current and charge density
matrix are derived and explained in a hopefully clear way. 
\end{abstract}

\thispagestyle{plain}
\pagestyle{plain}

\section{Introduction}
Non equilibrium Green's function methods are regularly used to calculate current and charge densities
in nanoscale (both molecular and semiconductor) conductors under bias. An overview of the theory
of molecular electronics can be found in Ref. \cite{paulsson.crc02} and for semiconductor nanoscale devices
see Ref. \cite{avik.chap02}. 

The aim of this text is to provide some intuitive explanations of one particle Green's 
functions in a compact form together with derivations of the expressions for the current and
the density matrix. It is not intended as a complete stand-alone tutorial, but rather as a
complement to Ref. \cite{paulsson.crc02,avik.chap02,zahid.mark02,datta.meso95}.

\section{Green's functions}
Discrete Schr\"odinger equation:
\beq
H \ket{n} = E \ket{n}
\eeq
We divide the Hamiltonian and wavefunction of the system into contact ($H_{1,2}$, $\ket{\psi_{1,2}}$) 
and device ($H_d$, $\ket{\psi_d}$) subspaces:
\beq
\mymat{3}{H_1 & \tau_1 & 0 \\ \tau_1^\dagger & H_d & \tau_2^\dagger \\ 0& \tau_2 & H_2}
\myvec{\ket{\psi_1}\\\ket{\psi_d}\\\ket{\psi_2}} =
E \myvec{\ket{\psi_1}\\\ket{\psi_d}\\\ket{\psi_2}} \label{eq.schrodinger}
\eeq
where $\tau_{1,2}$ describes the interaction between device and contacts.
In general we have $N$ contacts ($H_{1, \cdots, N}$) connecting ($\tau_{1, \cdots, N}$) the device $H_d$
to the reservoirs. Here we will assume that the contacts are independent, i.e., there are no cross terms ($\tau$)
between the different contacts.

We define the Green's function\footnote{Others may (and do) use the opposite sign.}:
\beq
\mybpar{E-H} G(E) =I
\eeq

\subsection{Why do we want to calculate the Green's function?}
The Green's function gives the response of a system to
a constant perturbation $\ket{v}$ in the Schr\"odinger equation:
\beq
H \ket{\psi} = E \ket{\psi} + \ket{v}
\eeq
The response to this perturbation is:
\beqa
\mybpar{E-H} \ket{\psi}&=& - \ket{v}  \hspace{2 cm}\rightarrow \\
\ket{\psi}&=&-G(E) \ket{v}
\eeqa

Why do we need the response to this type of perturbation?
Well, it turns out that it's usually easier (see next section) to calculate the Green's function than 
solve the whole eigenvalue problem\footnote{Especially for infinite systems.} 
and most (all for the one-particle system) properties  
of the system can be calculated from the Green's function. E.g.,
the wavefunction of the contact ($\ket{\psi_2}$) 
can be calculated if we know the  wavefunction on the device ($\ket{\psi_d}$). From third row of Eq.~\ref{eq.schrodinger}:
\beqa
H_2 \ket{\psi_2} + \tau_2 \ket{\psi_d} &=&E \ket{\psi_2} \hspace{2 cm} \rightarrow \\
\mybpar{E-H_2} \ket{\psi_2}&=&\tau_2 \ket{\psi_d} \hspace{2cm} \rightarrow \\
\mypsi{2}&=& g_2(E)  \tau_2 \mypsi{d} \label{eq.psi2}
\eeqa
where $g_2$ is the Green's function of the isolated contact 2 ($(E-H_2) g_2=I$). 

It is important to note that 
since we have an infinite system, we obtain two types of solutions for the Green's functions\footnote{
When the energy coincides with energy band of the contacts there are two solutions corresponding to
outgoing or incoming waves in the contacts. \label{fot.imaginary}
}, 
the retarded and the advanced\footnote{In practice these two solutions are usually obtained by adding
an imaginary part to the energy. By taking the limit to zero of the imaginary part one of the two solutions is obtained.
If the limit $\rightarrow 0^+$ is taken the retarded solution is found, $\rightarrow 0^-$ gives the advanced. 
This can be seen from the Fourier transform of the time dependent Green's function.} solutions corresponding to
outgoing and incoming waves in the contacts. 

\emph{Notation:}
We will denote the retarded Green's function with $G$ and the advanced with $G^\dagger$ 
(and maybe $G^R$ and $G^A$ occasionally).
Here, CAPITAL $G$ denotes the full Green's function and 
its sub-matrices $G_1$, $G_d$, $G_{1d}$ etc.
Lowercase is used for the Green's functions of the isolated subsystems, e.g., $\mybpar{E-H_2} g_2=I$.

Note that by using the retarded Green's function of the isolated contact ($g_2$) in Eq.~\ref{eq.psi2} 
we obtain the solution corresponding 
to a outgoing wave in the contact. Using the advanced Green's function ($g_2^\dagger$) would give the solution 
corresponding to an incoming wave.

\subsection{Self-Energy}
The reason for calculating the Green's function is that it is easier that solving the Schr\"odinger equation. 
Also, the Green's function of the device ($G_d$) can be calculated separately without calculating the whole Green's function ($G$).
From the definition of the Green's function we obtain:
\beq
\mymat{3}{E-H_1 & -\tau_1 & 0 \\ -\tau_1^\dagger & E-H_d & -\tau_2^\dagger \\ 0& -\tau_2 & E-H_2}
\mymat{3}{G_1 & G_{1d} & G_{12} \\G_{d1} & G_{d} & G_{d2} \\G_{21} & G_{2d} & G_{2}} =
\mymat{3}{I &0&0\\0&I&0\\0&0&I}
\eeq
Selecting the three equations in the second column:
\beqa
\mybpar{E-H_1} G_{1d} -\tau_1 G_{d}&=&0\label{eq.tmp1}\\
-\tau^\dagger_1 G_{1d}+\mybpar{E-H_d} G_d -\tau^\dagger_2 G_{2d}&=&I \label{eq.tmp2}\\
\mybpar{E-H_2} G_{2d} -\tau_2 G_{d}&=&0\label{eq.tmp3}
\eeqa
We can solve Eqs.~\ref{eq.tmp1} and \ref{eq.tmp3} for $G_{1d}$ and $G_{2d}$:
\beqa
G_{1d}&=&g_1 \tau_1 G_{d}\label{eq.g1d} \\
G_{2d}&=&g_2 \tau_2 G_{d} \label{eq.g2d}
\eeqa
substitution into Eq.~\ref{eq.tmp2} gives:
\beq
-\tau^\dagger_1 g_1 \tau_1 G_{d}+\mybpar{E-H_d} G_d -\tau^\dagger_2 g_2 \tau_2 G_{d}=I \\
\eeq
from which $G_d$ is simple to find:
\beq
G_d=\mybpar{E-H_d-\Sigma_1 -\Sigma_2}^{-1}
\eeq
where $\Sigma_1=\tau^\dagger_1 g_1 \tau_1$ and $\Sigma_2=\tau^\dagger_2 g_2 \tau_2$ are the so called self-energies.

Loosely one can say that the effect of the contacts on the device is to add the self-energies to the device Hamiltonian
since when we calculate the Green's function on the device we just calculate the 
Green's function for the effective Hamiltonian
$H_{\mbox{\tiny effective}}=H_d+\Sigma_1+\Sigma_2$. However, we should keep in mind that we can 
only do this when we calculate the 
Green's function. The eigen-values and -vectors of this effective Hamiltonian are not 
quantities we can interpret easily. 

For ``normal'' contacts, the surface Green's functions $g_1$ and $g_2$ used to calculate the self-energies 
are usually calculated using the periodicity of the contacts, this method is described in detail in appendix B 
of Ref. \cite{zahid.mark02} and in section 3 of Ref. \cite{avik.chap02}.

\subsection{The spectral function}
Another important use of the Green's function is the spectral function:
\beq
A= i\mybpar{G-G^\dagger}
\eeq
which gives the DOS and \emph{all} solutions to the Schr\"odinger equation!

To see this we first note that for any perturbation $\ket{v}$ we get two solutions ($\ket{\psi^R}$ and $\ket{\psi^A}$) 
to the perturbed Schr\"odinger equation:
\beq
(E-H) \mypsi{} = -\ket{v}
\eeq
from the advanced and retarded Green's functions:
\beqa
\ket{\psi}^R&=&-G \ket{v} \\
\ket{\psi}^A&=&-G^\dagger \ket{v}
\eeqa
The difference of these solutions ($\ket{\psi^R}-\ket{\psi^A}=(G-G^\dagger) \ket{v}=-i A \ket{v}$) 
is a solution to the Schr\"odinger equation:
\beq
(E-H)(\ket{\psi^R}-\ket{\psi^A})=(E-H)(G-G^\dagger) \ket{v}=(I-I) \ket{v}=0
\eeq
which means that $\ket{\psi}= A \ket{v}$ is a solution to the Schr\"odinger equation for any vector $\ket{v}$!

To show that the spectral function actually gives \emph{all} solutions to the Schr\"odinger equation 
is a little bit more complicated and we
need the expansion of the Green's function in the eigenbasis:
\beq
G=\frac{1}{E+i \delta-H}=\mysum{k}{}{\frac{\ketbra{k}{k}}{E+i \delta-\epsilon_k}}
\eeq
where the $\delta$ is the small imaginary part (see footnote \ref{fot.imaginary}),
$\ket{k}$'s are all eigenvectors\footnote{Normalized!} to $H$ with the corresponding 
eigenvalues $\epsilon_k$. Expanding the spectral function in the eigenbasis gives:
\beqa
A&=&i \mybpar{\frac{1}{E+i \delta -H}-\frac{1}{E-i \delta -H}} \\
&=&i\mysum{k}{}  {\ketbra{k}{k}\mybpar{\frac{1}{E+i \delta-\epsilon_k}-\frac{1}{E-i \delta-\epsilon_k}}}\\
&=&\mysum{k}{} {\ketbra{k}{k}\frac{2 \delta}{\mybpar{E-\epsilon_k}^2 + \delta^2}}
\eeqa
where $\delta$ is our infinitesimal imaginary part of the energy.
Letting $\delta$ go to zero gives:
\beqa
A&=&2 \pi \mysum{k}{} {\delta\mybpar{E-\epsilon_k} \ketbra{k}{k}} \label{eq.Aeig}
\eeqa
(here $\delta(E-\epsilon_k)$ is the delta function) which can be seen since 
$\frac{2 \delta}{\mybpar{E-\epsilon_k}^2 + \delta^2}$ goes to zero everywhere but
at $E=\epsilon_k$, integrating over $E$ (with a test function) gives the $2 \pi \delta\mybpar{E-\epsilon_k}$
factor. Eq.~\ref{eq.Aeig} shows that the spectral 
function gives us all solutions to the Schr\"odinger equation.

\section{Response to an incoming wave}
In the non-equilibrium case, reservoirs with different chemical potentials will inject electrons and
occupy the states corresponding to incoming waves in the contacts. Therefore, we want to 
find the solutions corresponding to these incoming waves.
 
Consider contact $1$ isolated from the other contacts and the device. 
At a certain energy we have solutions corresponding to an incoming wave that is totally reflected at
the end of the contact. We will denote these solutions with $\mypsi{1,n}$ where $1$ is the contact number
and $n$ is a quantum number (we may have several modes in the contacts). We can find all these solutions
from the spectral function $a_1$ of the isolated contact (as described above).

Connecting the contacts to the device we can calculate the wavefunction on the whole system caused 
by the incoming wave in contact $1$. To do this we note that a wavefunction should be of the form 
$\mypsi{1,n}+\ket{\psi^{R}}$ where  $\mypsi{1,n}$ is the totally reflected wave and $\ket{\psi^R}$ is the 
retarded response of the whole system.
Putting in the anzats $\mypsi{1,n}+\ket{\psi^R}$ into the Schr\"odinger equation gives:
\beqa
\myvec{H_1 + \tau_1 +\\ H_d + \tau_1^\dagger +\tau_2^\dagger +\\H_2+\tau_2} 
\mybpar{\mypsi{1,n}+\ket{\psi^R}}&=&
E  \mybpar{\mypsi{1,n}+\ket{\psi^R}} \\
\myvec{E \mypsi{1,n}+\\ \tau_1^\dagger \mypsi{1,n}+\\0} +
\myvec{H_1 +\tau_1 +\\ H_d + \tau_1^\dagger +\tau_2^\dagger +\\H_2+\tau_2} \ket{\psi^R}&=&
E  \mybpar{\mypsi{1,n}+\ket{\psi^R}} \\
H \ket{\psi^R} &=& E \ket{\psi^R} -\tau_1^\dagger \mypsi{1,n}
\eeqa
(note the slight change in notation)
and we see that $\ket{\psi^R}$ is noting else the response of the whole system to 
a perturbation of $-\tau_1^\dagger \mypsi{1,n}$, c.f., Eq.~\ref{eq.psi2}:
\beq
\ket{\psi^R}=G \tau_1^\dagger \mypsi{1,n} \label{eq.scatteringstates}
\eeq
It is important to realize that the scattering states generated from Eq.~\ref{eq.scatteringstates},
using all possible incoming waves from each contact, form a complete\footnote{Except for localized 
states in the device region.} ON set of solutions to the full Schr\"odinger equation\cite{ballentine}.
Note that we have chosen the retarded response which means that the only part of 
the wave that is traveling towards the 
device is the incoming wave (part of $\mypsi{1,n}$). We will make full use of this fact below.

It will be useful to have the expressions for the device wavefunction $\mypsi{d}$ and contact wavefunction 
($\mypsi{1,2}$). The device part is straightforward:
\beq
\mypsi{d}=G_d \tau_1^\dagger \mypsi{1,n} \label{eq.psid}
\eeq
and from Eq.~\ref{eq.psi2} or Eq.~\ref{eq.g2d}:
\beq
\mypsi{2}= g_2 \tau_2 \mypsi{d}=g_2 \tau_2 G_d \tau_1^\dagger \mypsi{1,n} \label{eq.psi2b}
\eeq
Note, to calculate the wavefunction in the contact containing the incoming wave (contact $1$) 
we need to add the incoming wave, giving a slightly more complicated expression:  
\beq
\mypsi{1}= \mybpar{1+g_1 \tau_1 G_d \tau_1^\dagger} \mypsi{1,n} \label{eq.psi1}
\eeq

Knowing the wavefunctions corresponding to incoming waves in different contacts enables us
to fill up the different solutions according to the electron reservoirs filling the contacts.

\section{Charge density matrix}
In the non equilibrium case we are often interested in two quantities: the current and the 
charge density matrix. Lets start with the charge density (which allows us to use a self-consistent
scheme to describe charging).
 
The charge density matrix is defined as:
\beq
\rho=\mysum{k}{} f(k,\mu) \ket{\psi_k} \bra{\psi_k} 
\eeq
where the sum runs over all states with the occupation number $f(E_k,\mu)$
(pure density matrix) (note the similarity with the spectral function $A$, in equilibrium you find the density
matrix from $A$ and not as described below).
In our case, the occupation number is determined by the reservoirs filling the incoming waves in the contacts
such that:
\beq
f(E_k,\mu_1)=\frac{1}{1+e^{(E_k-\mu_1)/{k_B T}}}
\eeq
is the Fermi-Dirac function with the chemical potential ($\mu_1$) 
and temperature ($T$) of the reservoir responsible for injecting the 
electrons into the contacts.

The wavefunction on the device given by an incoming wave in contact $1$ (see Eq.~\ref{eq.psid}) is:
\beq
\mypsi{d,k}=G_d \tau_1^\dagger \mypsi{1,k}
\eeq
Adding up all states from contact $1$ gives:
\beqa
 \rho_d[\mbox{contact $1$}]&=&
  \myint{E=-\infty}{\infty}  \mbox{d}E \; \mysum{k}{} f(E,\mu_1)  \delta(E-E_k) \ket{\psi_{d,k}}\bra{\psi_{d,k}} \\
 &=&\myint{E=-\infty}{\infty}  \mbox{d}E \; f(E,\mu_1) \mysum{k}{}  \delta(E-E_k)
G_d \tau_1^\dagger \ket{\psi_{1,k}} \bra{\psi_{1,k}} \tau_1 G^\dagger_d  \\
 &=&\myint{E=-\infty}{\infty}  \mbox{d}E \;f(E,\mu_1) 
G_d \tau_1^\dagger \mybpar{\mysum{k}{} \delta(E-E_k)  
  \ket{\psi_{1,k}} \bra{\psi_{1,k}}} \tau_1 G^\dagger_d \\
&=&\left[\mbox{\protect{Eq.~\ref{eq.Aeig}}}\right] =\myint{E=-\infty}{\infty} \mbox{d}E \; f(E,\mu_1) 
G_d \tau_1^\dagger \frac{a_1}{2 \pi} \tau_1 G^\dagger_d 
\eeqa
introducing the new quantity $\Gamma_1=\tau_1^\dagger a_1 \tau_1=i\mybpar{\Sigma_1-\Sigma_1^\dagger}$ 
we obtain the simple formula:
\beq
\rho[\mbox{from contact $1$}]= \frac{1}{2\pi}\myint{E=-\infty}{\infty} \mbox{d}E \; 
f(E,\mu_1) G_d \Gamma_1 G^\dagger_d 
\eeq

The total charge density thus becomes a sum over all contacts:
\beq
\rho= \frac{2 \; \mbox{(for spin)}}{2\pi} \myint{E=-\infty}{\infty}  \mbox{d}E \;\mysum{i}{} 
f(E,\mu_i) G_d \Gamma_i G^\dagger_d 
\eeq

\section{Probability Current}
Having different chemical potentials in the reservoirs filling the contacts gives rise to a current.
In the next section we will calculate this current in a similar way as the charge density was calculated.
But to do this we need an expression for the current from the wavefunction.

In the continuum case we can calculate the current from the velocity operator. However, for a discrete
Hamiltonian it is not so clear what the velocity operator is. Therefore, we derive an expression for the 
current from the continuity equation (using two contacts). In steady-state, the probability to find an 
electron on the device ($\mysum{i}{} |\psi_i|^2$ where 
the sum runs over the device subspace) is conserved:
\beqa
0&=&\mydt{\mysum{i}{} |\psi_i|^2}=\mysum{i}{} \mydt{ \braket{\psi}{i}\braket{i}{\psi}}=\mysum{i}{} \mybpar{\mydt{\braket{\psi}{i}} \braket{i}{\psi}+ \braket{\psi}{i} \mydt{\braket{i}{\psi}}}\\
&=&\frac{i}{\hbar} \mysum{i}{} \mybpar{ \brahket{\psi}{H}{i} \braket{i}{\psi}- \braket{\psi}{i} \brahket{i}{H}{\psi}}=\frac{i}{\hbar} \mybpar{ \brahket{\psi}{H}{\psi_d}- \brahket{\psi_d}{H}{\psi}}\\
&=&\frac{i}{\hbar} \mybpar{ \brahket{\psi}{H_d+\tau_1+\tau_2}{\psi_d}- 
	\brahket{\psi_d}{H_d+\tau_1^\dagger+\tau_2^\dagger}{\psi}}\\
&=&\frac{i}{\hbar} \mybpar{ \left[\brahket{\psi_1}{\tau_1}{\psi_d}-\brahket{\psi_d}{\tau_1^\dagger}{\psi_1}\right]+ 
	\left[\brahket{\psi_2}{\tau_2}{\psi_d}-\brahket{\psi_d}{\tau_2^\dagger}{\psi_2}\right]}
\eeqa
We interpret the term in the first (square) bracket as the incoming probability current into the device from contact $1$
and the second bracket from contact $2$. Generalizing to an arbitrary contact $j$ gives us the electric current (at
one energy) as 
the charge (-e) times the probability current:
\beq
i_j=-\frac{i e}{\hbar} \mybpar{ \brahket{\psi_j}{\tau_j}{\psi_d}-\brahket{\psi_d}{\tau_j^\dagger}{\psi_j}}
\eeq
where $i_j$ is defined as positive for a current from the contacts into the device.
We can now put in the expressions for the wavefunctions in the same way as for the density matrix. 
 
\section{Electrical Current}
To calculate the total current through the device we only need to put in the wavefunction of the 
device and the contacts ($\mypsi{d}$, $\mypsi{1}$, $\mypsi{2}$)
from Eqs.~\ref{eq.psid}, \ref{eq.psi1} and \ref{eq.psi2} and add all the contributions together.
Thus the current into the device from a incoming wave of one energy ($E$) 
in contact $1$ ($\mypsi{1,n}$) through the coupling defined by $\tau_2$ is:
\beqa
i_{\mbox{\tiny 2 from 1}}&=&-\frac{i e}{\hbar} 
\mybpar{\brahket{\psi_2}{\tau_2}{\psi_d}-\brahket{\psi_d}{\tau_2^\dagger}{\psi_2}}\\
&=&-\frac{i e}{\hbar} (\brahket{\psi_{1,n}}{
   \tau_1 G^\dagger_d \tau_2^\dagger g_2^\dagger  \tau_2 G_d \tau_1^\dagger}{\psi_{1,n}}-  \brahket{\psi_{1,n}}{
  \tau_1 G^\dagger_d \tau_2^\dagger g_2 \tau_2 G_d \tau_1^\dagger   }{\psi_{1,n}})\\
&=&-\frac{i e}{\hbar} \brahket{\psi_{1,n}}{ \tau_1 G^\dagger_d \tau_2^\dagger \mybpar{g_2^\dagger-g_2}  
\tau_2 G_d \tau_1^\dagger}{\psi_{1,n}} ) \\
&=&\frac{e}{\hbar}
\brahket{\psi_{1,n}}{ \tau_1 G^\dagger_d \Gamma_2 G_d \tau_1^\dagger}{\psi_{1,n}} 
\eeqa 
Adding over the modes $n$ and noting that the levels are filled from the reservoir 
connected to contact $1$ gives (2 for spin):
\beqa
I_{\mbox{\tiny 2 from 1}}&=&2 \frac{e}{\hbar} \myint{E=-\infty}{\infty}  \mbox{d}E \,f(E,\mu_1) \mysum{n}{}
\delta(E-E_n) \brahket{\psi_{1,n}}{ \tau_1 G^\dagger_d \Gamma_2 G_d \tau_1^\dagger}{\psi_{1,n}}\\
&=&\frac{2 e}{\hbar} \myint{E=-\infty}{\infty}  \mbox{d}E \, f(E,\mu_1) \mysum{m,n}{} \delta(E-E_n) 
\brahket{\psi_{1,n}}{ \tau_1}{m} \brahket{m}{G^\dagger_d \Gamma_2 G_d \tau_1^\dagger}{\psi_{1,n}}\\
&=&\frac{2 e}{\hbar} \myint{E=-\infty}{\infty} f(E,\mu_1) \mysum{m}{}
\bra{m}{G^\dagger_d \Gamma_2 G_d } \tau_1^\dagger \mybpar{ \mysum{n}{} \delta(E-E_n) \ket{\psi_{1,n}}\bra{\psi_{1,n}} }\tau_1\ket{m} \\
&=&\frac{2 e}{\hbar} \myint{E=-\infty}{\infty} \mbox{d}E \, f(E,\mu_1) \mysum{m}{}
\brahket{m}{G^\dagger_d \Gamma_2 G_d \tau_1^\dagger \frac{a_1}{2 \pi} \tau_1}{m} \\
&=&\frac{e}{\pi \hbar} \myint{E=-\infty}{\infty} \mbox{d}E \, f(E,\mu_1) \mbox{Tr} \mybpar{
{G^\dagger_d \Gamma_2 G_d \Gamma_1}}
\eeqa

To get the total current through the device the current from contact two have to be subtracted away:
\beq
I=\frac{e}{\pi \hbar} \myint{E=-\infty}{\infty} \mbox{d}E \mybpar{f(E,\mu_1)-f(E,\mu_2)} \mbox{Tr} \mybpar{
{G^\dagger_d \Gamma_2 G_d \Gamma_1}}
\eeq
which is exactly the Landauer formula for the current.

\section*{Acknowledgments}
The author is grateful for the inspiration from Supriyo Datta and 
Mads Brandbyge. The revised version was initiated from discussions 
with Casper Krag and Carsten Rostg\aa rd. 

\bibliography{NEGF}
\bibliographystyle{unsrt}

\end{document}